\documentclass[9pt,twocolumn,twoside]{osajnl}

\journal{ol} 

\setboolean{shortarticle}{false} 

\usepackage{braket}
\usepackage{microtype} 
\usepackage{pgfplots}
\usepackage{gensymb}
\usepackage{amsmath}
\usepackage{amsfonts}
\usepackage{amssymb}
\usepackage{bm}
\usepackage{mathtools} 
\usepackage{braket}
\usepackage{comment}
\usepackage{subfig}

\newcommand{\MC}[1]{\textcolor{orange}{#1}}

\usepgfplotslibrary{colorbrewer}
\pgfplotsset{cycle list/Dark2-6}

\ifthenelse{\boolean{shortarticle}}{\colorlet{color2}{color2b}}{\colorlet{color2}{color2}} 

\title{Spatial Mode Diversity for Robust Free-Space Optical Communications}

\author[1,2,*]{Mitchell A. Cox}
\author[2]{Carmelo Rosales-Guzm\'an}
\author[1]{Ling Cheng}
\author[2]{Andrew Forbes}

\affil[1]{School of Electrical and Information Engineering, University of the Witwatersrand, Johannesburg 2050, South Africa}
\affil[2]{School of Physics, University of the Witwatersrand, Johannesburg 2050, South Africa}

\affil[*]{Corresponding author: mitchell.cox@wits.ac.za}

\dates{Compiled \today}

\ociscodes{(060.2605) Free-space optical communication; (010.1300) Atmospheric propagation; (030.7060) Turbulence.}

\doi{\url{http://dx.doi.org/10.1364/ol.XX.XXXXXX}}

\begin{abstract}
Free-space communication links are severely affected by atmospheric turbulence, which causes degradation in the transmitted signal.  One of the most common solutions to overcome this is to exploit diversity. In this approach, information is sent in parallel using two or more transmitters that are spatially separated, with each beam therefore experiencing different atmospheric turbulence, lowering the probability of a receive error. In this work we propose and experimentally demonstrate a generalization of diversity based on spatial modes of light, which we have termed \textit{modal diversity}. We remove the need for a physical separation of the transmitters by exploiting the fact that spatial modes of light experience different perturbations, even when travelling along the same path. For this proof-of-principle we selected modes from the Hermite-Gaussian and Laguerre-Gaussian basis sets and demonstrate an improvement in Bit Error Rate by up to 54\%. We outline that modal diversity enables physically compact and longer distance free space optical links without increasing the total transmit power.
\end{abstract}

\setboolean{displaycopyright}{false}

\begin{document}

\maketitle
\thispagestyle{fancy}

\ifthenelse{\boolean{shortarticle}}{\ifthenelse{\boolean{singlecolumn}}{\abscontentformatted}{\abscontent}}{}


\noindent Free Space Optical (FSO) communication has been the subject of significant research in recent years due to the impending capacity crunch \cite{Richardson1}. Using FSO communications between buildings and even for Earth to satellite communications is of primary interest because of ultra high bandwidth capabilities when technologies such as Mode Division Multiplexing (MDM) are employed \cite{GibsonII2004, Wang2012}. Here, several orthogonal modes are used, each carrying a separate data stream to multiply the overall data rate of a system. Historically only the azimuthal subset of the Laguerre-Gauss (LG$_\ell^p$) basis has been utilised, specified by mode index $\ell$, which relates to the orbital angular momentum of each photon, $\ell\hbar$. The use of the radial LG index, $p$, is uncommon but is required to harness the full capacity of the basis \cite{Zhao2015,Trichili2016}. The Hermite-Gauss (HG$_n^m$) basis can also be used for MDM and is specified by mode indices $n$ and $m$, and beyond scalar modes, Cylindrical Vector Vortex (CVV) modes have also been the subject of attention for optical communications in both free space and optical fibres \cite{Milione2015,Cox2016,ndagano2017b}.

Upon propagation through the atmosphere, turbulence distorts the wave-front, amplitude and phase of the launched modes, called fading. Fading leads to errors in a communication system with a certain probability, and hence deteriorates the overall performance. In an MDM system this fading may also manifest as crosstalk between channels. Fading is obviously dependent on the path that the beam propagates through and so when several independent paths are used, typically with a separation of at least $r_0$ (the Fried Parameter, which is a measure of the transverse distance over which the refractive index is correlated \cite{Fried1965}), the probability of error is typically reduced \cite{Navidpour2007a}. This is the basis for what is known as diversity and the so-called diversity gain can often be predicted by using probabilistic models applied to a range of mode types \cite{Churnside1987b,Kashani2015,Anguita2009,Zhou2015,Aksenov2016,Ren2015,Zou2017}.  Diversity can be considered the complement to MDM, reducing errors rather than increasing capacity.  Yet while MDM has been extensively studied, the specific use of optical modes for diversity has not been studied before.

In this letter, we break the paradigm that diversity requires beams to propagate along differing paths in space.  Instead, we demonstrate that modes of different spatial profiles are perturbed differently by atmospheric turbulence, and hence carefully chosen modes will result in a diversity gain even if they propagate on the same path, a concept we refer to as \textit{modal diversity}.  Inspired by recent demonstrations of the robustness of HG modes in turbulence \cite{ndagano2017}, we provide an experimental demonstration of atmospheric turbulence mitigation using modal diversity with orthogonal LG and HG co-propagating modes without $r_0$ aperture separation. With this novel diversity method, the resilience of free-space optical links can be engineered by a judicious choice of the spatial modes, enabling more reliable and longer distance communications, addressing a major drawback of FSO communications when compared to radio. 

In a typical diversity model, multiple transmitters (lasers) transmit identical signals, $x_i(t)$, at the same time with intensity $g_i$. These signals propagate through separate channels represented by channel impulse responses, $h_i(t)$. The signal is then detected by a single receiver (photodiode) with receiver sensitivity, $r$. The resulting received signal, $y(t)$ is then found from 

\begin{equation}
\label{eq:model}
y(t) = r \sum_i g_i x_i(t) * h_i(t) + n(t).
\end{equation}

\noindent where an Additive White Gaussian Noise (AWGN) component, $n(t)$, may be incorporated.  Traditionally, ``separate channels'' means separate paths, which in turbulence means a physical separation of at least $r_0$.  Now we will alter this paradigm to interpret ``separate channels'' as distinct modes with differing behaviour in turbulence. 

If the signal is strong then the contribution of noise to errors is insignificant and the dominant cause of errors is the channel gain itself. If the channel gains are statistically independent then we can write the overall probability of an error in the diversity case as

\begin{equation}
\mathrm{Pr}[E_{\mathrm{diversity}}] = \prod_i \mathrm{Pr}[E_{i}],
\end{equation}

\noindent where the probability of an error occurring for the $i^{th}$ channel is defined as $\mathrm{Pr}[E_i]$.  This always results in a lower probability of error than a single channel case (since $\mathrm{Pr}[E_i] \leq 1$). This standard diversity scheme is similar to Equal Gain Combining (EGC) and will only show a diversity gain if the individual channels are statistically independent \cite{Brennan1959}.

\medskip

To implement the above using modal diversity, identical order yet orthogonal HG$_n^m$ and LG$_\ell^p$ beams were chosen, motivated by the fact that HG modes are robust to tip/tilt which are the primary aberrations of atmospheric turbulence \cite{ndagano2017}. The order of the beams is given by $N = n + m = 2p + |\ell|$ for HG and LG beams respectively. The completeness property of both bases allow us to express any element of one basis as a linear combination of elements from the other basis using the transformation relations below \cite{ONeil2000}:

\begin{equation}
\label{eq:LGtransform}
LG_{n,m}(x,y,z)=\sum_{k=0}^N i^kb(n,m,k)HG_{N-k,k}(x,y,z)
\end{equation}
\begin{equation}
\label{eq:LGtransformB}
b(n,m,k)=\left[\frac{(N-k)!k!}{2^Nn!m!}\right]^{1/2}\frac{1}{k}\frac{d^k}{dt^k}[(1-t)^n(1+t)^m]|_{t=0}
\end{equation}

The $LG$ modes have been written in terms of $n$ and $m$, which are indices typically used for HG modes. Traditionally, $LG$ modes are given in terms of azimuthal index $\ell$ and radial index $p$, which can be recovered as $\ell=n-m$ and $p=\mbox{min}(n,m)$. For example, the LG$_2^1$ mode may be written as

\begin{equation}
LG_2^1 = \frac{1}{2}HG_4^0 - \frac{i}{2}HG_3^1 + 0 \times HG_2^2 - \frac{i}{2}HG_1^3 - \frac{1}{2}HG_0^4.
\end{equation}

\begin{figure}[h]
\centering
\begin{tikzpicture}
    \begin{axis}[
    	title={$HG_2^2$ Beam},
        width=2.4cm,
        height=2.4cm,
        scale only axis,
        enlargelimits=false,
        axis on top,
        ticklabel style = {font=\small},
        xtick={25,75},
        ytick={25,75},
        ytick style={draw=none},
        xtick style={draw=none},
        ylabel={Strehl Ratio},
    	xticklabels={$HG^2_2$,$LG^1_2$},
        yticklabels={0.95,1.0},
    ]
      \addplot graphics[xmin=0,xmax=100,ymin=0,ymax=100] {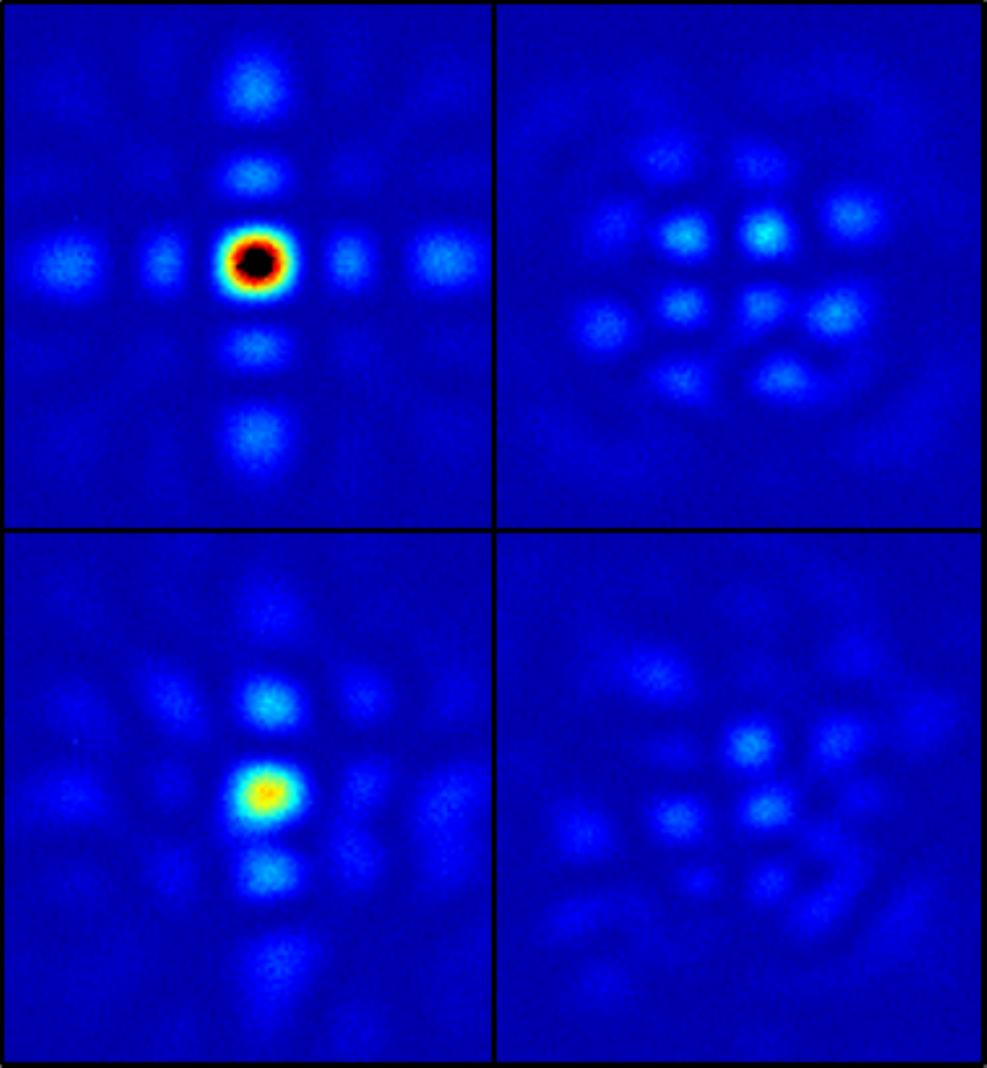};
    \end{axis}
  \end{tikzpicture}%
  \qquad%
  \begin{tikzpicture}
    \begin{axis}[
        title={$LG_2^1$ Beam},
        colormap/jet, 
		colorbar,
        colorbar style={ytick={0.0,0.5,1.0},yticklabel style={
            /pgf/number format/.cd,
                fixed,
                fixed zerofill, precision=1
        }},
        width=2.4cm,
        height=2.4cm,
        scale only axis,
        enlargelimits=false,
        axis on top,
        ticklabel style = {font=\small},
        xtick={25,75},
        ytick={25,75},
        ytick style={draw=none},
        xtick style={draw=none},
    	xticklabels={$LG^1_2$,$HG^2_2$},
        yticklabels={,}
    ]
      \addplot graphics[xmin=0,xmax=100,ymin=0,ymax=100] {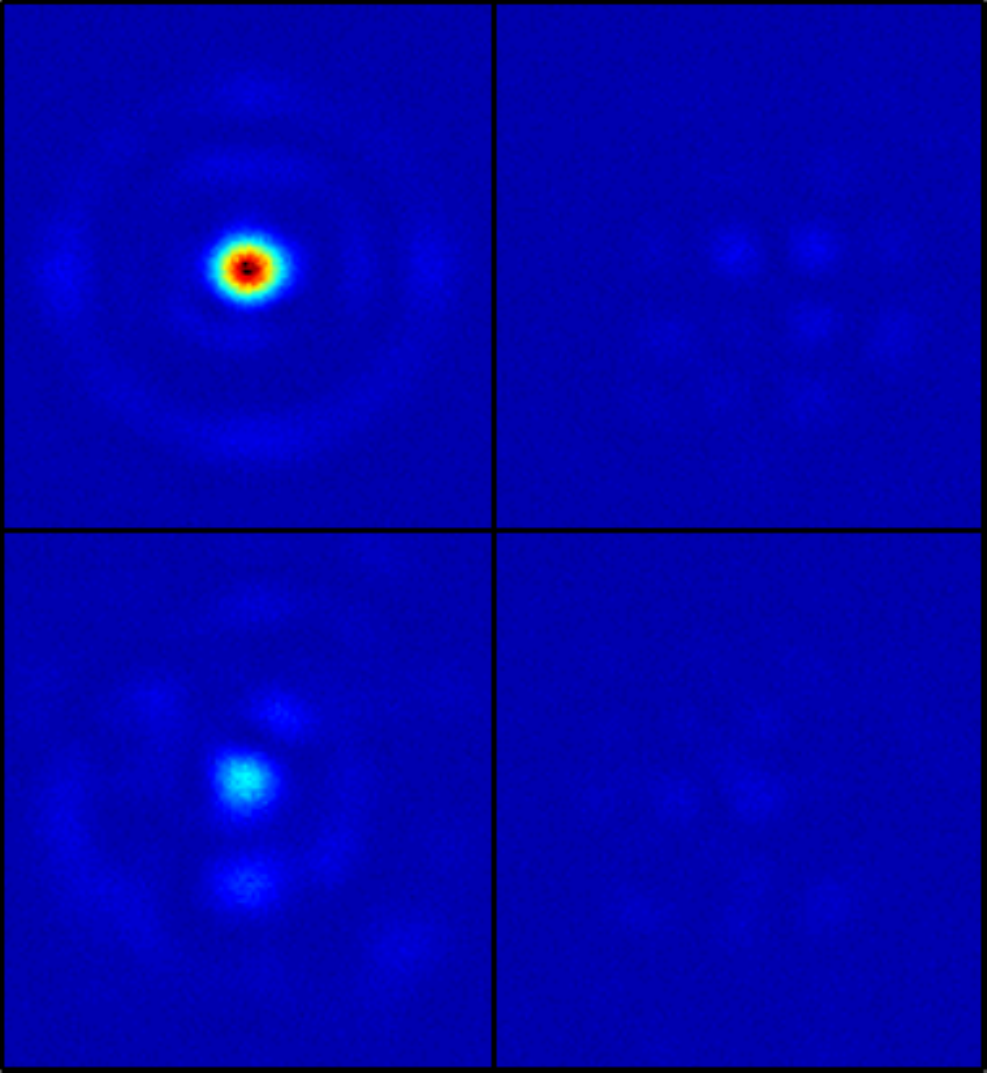};
    \end{axis}
  \end{tikzpicture}  

  \caption{\label{fig:crosstalk2x2} Experimental demonstration of the orthogonality of HG$_2^2$ and LG$_2^1$ modes with no turbulence (SR=1.0) and slight turbulence (SR=0.95). On the left, only an HG$_2^2$ is transmitted and it is clear that there is no LG$_2^1$ component after modal decomposition. Similarly, on the right there is an LG$_2^1$ component but no HG$_2^2$ component in the decomposed beam. }
  \end{figure}

Notice that the HG$_2^2$ component has a zero weighting, making the HG$_2^2$ mode orthogonal to the LG$_2^1$ mode. Figure~\ref{fig:crosstalk2x2} shows an experimental test of this orthogonality. Conveniently, both of these modes have an order of $N=4$, however, not all LG modes contain an orthogonal HG mode with the same mode order, and vice versa. Similarly, HG$_4^4$ is orthogonal to LG$_6^1$ with $N=8$. Both of these mode sets were tested in the experiment.  

\begin{figure*}[ht]
  \centering
  \includegraphics[width=1\textwidth]{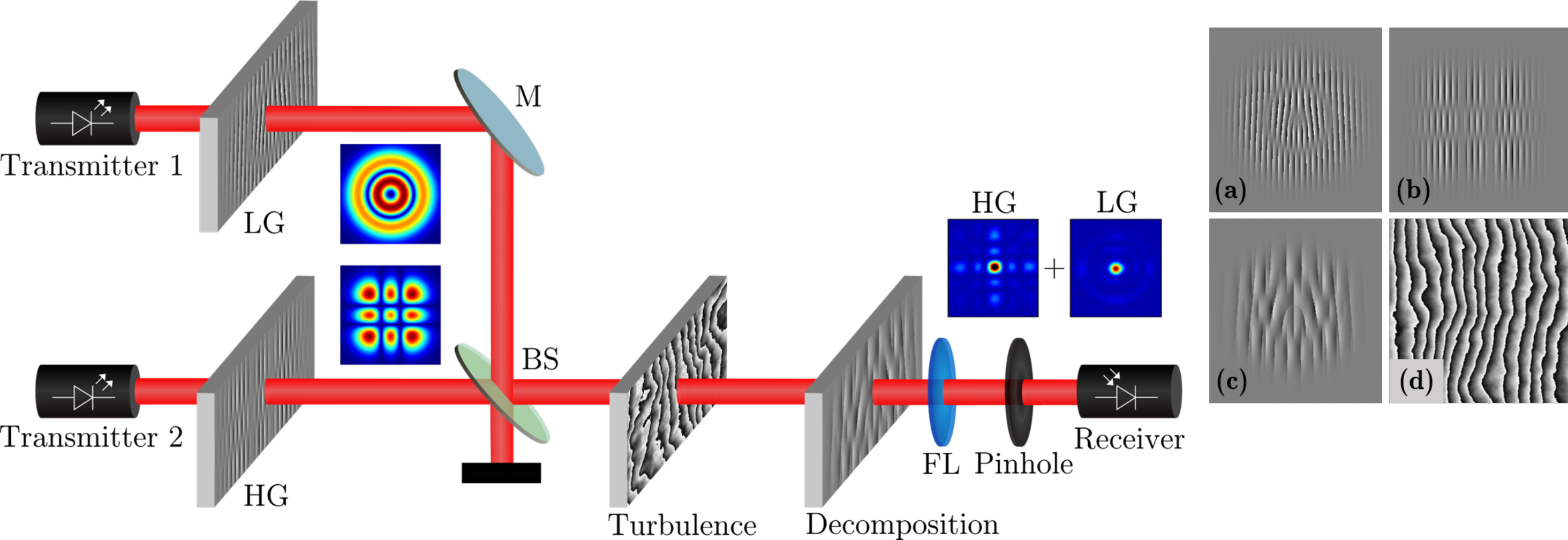}
\caption{\label{fig:setup}Experimental setup showing two transmitters and a single receiver for transmit diversity. The insets to the right are holograms for (a) generating an LG$_2^1$ beam, (b) generating an HG$_2^2$ beam, (c) superposition of the aforementioned holograms and (d) example turbulence with a grating with SR=0.7. The grey area on the holograms is due to complex amplitude modulation. }
\end{figure*}

The experimental setup consisted of two laser diodes which were individually modulated with identical On-Off Keying (OOK-NRZ) signals. The modulation was intentionally done at low bandwidth to avoid non-ideal variables related to high bandwidth communication. As shown in Fig.~\ref{fig:setup}, each beam was transformed using a Spatial Light Modulator (SLM) into either an LG or HG mode. The first diffraction order from each SLM was spatially filtered and both beams were then combined using a beam splitter to propagate co-linearly. Kolmogorov turbulence and modal decomposition were performed by another SLM \cite{Andrews2005,Rosales2017}.  Finally, the resulting beam was filtered using a $50~\mu$m precision pinhole after which the intensity was measured by a photodiode for demodulation. The experiment was performed with varying turbulence strengths corresponding to a range of atmospheric coherence lengths ($r_0$) from 0.1 to 17~mm. These coherence lengths correspond to the Strehl Ratios 0.01 to 0.9 when applied to the beams with mode order $N=4$. For each turbulence strength and each mode, one million random bits were tested against 1024 random Kolmogorov phase screens. The Bit Error Rate (BER) was then determined for each test case.  Each mode was first tested individually at a specific intensity in a conventional Single Input Single Output (SISO) configuration to determine baseline BER performance for the experimental setup without diversity. Transmit diversity was then tested by transmitting a random bit-stream over the HG and LG modes simultaneously. It is important to note that in the diversity case half of the SISO intensity was used for each transmitter, resulting in the same total intensity as in the SISO case at the receiver. This is done to ensure a fair Signal to Noise Ratio comparison to the SISO case. 

\begin{figure}[tbh]
	\centering
	\begin{tikzpicture}

	\begin{semilogyaxis}[ %
    compat=1.3,
	grid=major,
    smooth,
    thick,
	grid style={dashed,gray!20},
	width={0.48\textwidth},
	height=6cm,
	legend entries={$HG_2^2$, $LG_2^1$, Diversity},
	legend cell align=left,
	legend style={draw=none},
	legend pos = north east,
	xlabel={Turbulence $r_0$ [mm]},
    xticklabel style={/pgf/number format/.cd,fixed,fixed zerofill,precision=1},
    ylabel shift = -5 pt,
    scaled x ticks=false,
	minor x tick num=1,
	minor y tick num=1,
	xmin=0,
	xmax=17,
    ymax=1,
    ymin=0.01,
	ylabel={Bit Error Rate (BER)},
    axis line style=thick,
    cycle list name=Dark2-6
	] %
    
    
    \addplot+[dashed,line width = 1,mark=square*] table [x=r0, y=HG22, col sep=comma] {summary.csv};
    \addplot+[dashed,line width = 1,mark=*] table [x=r0, y=LG21, col sep=comma] {summary.csv};
    \addplot+[line width = 1,mark=diamond*] table [x=r0, y=EGC2221, col sep=comma] {summary.csv};
    
	
    \addplot+[solid,mark=none,color=black] coordinates {(1.4, 0.001) (1.4, 1)};
    
	\end{semilogyaxis} %
	\end{tikzpicture}
	
	\caption{\label{fig:ber4} Bit Error Rate of modes with order $N=4$ with varying turbulence strength showing a clear improvement for the diversity case (solid lines). The vertical line indicates the approximate size of the beams.}
\end{figure}
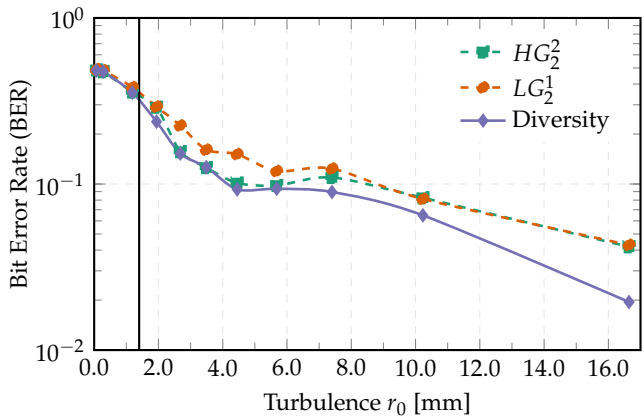

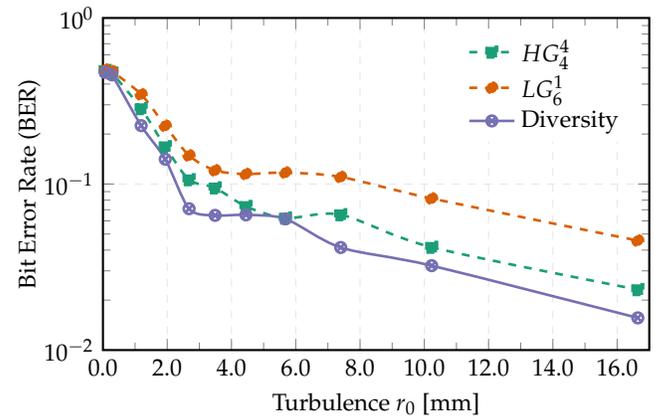
\begin{figure}[tbh]
	\centering
	\begin{tikzpicture}

	\begin{semilogyaxis}[ %
    compat=1.3,
	grid=major,
    smooth,
    thick,
	grid style={dashed,gray!20},
	width={0.48\textwidth},
	height=6cm,
	legend entries={$HG_4^4$, $LG_6^1$, Diversity},
	legend cell align=left,
	legend style={draw=none},
	legend pos = north east,
    ylabel shift = -5 pt,
	xlabel={Turbulence $r_0$ [mm]},
    xticklabel style={/pgf/number format/.cd,fixed,fixed zerofill,precision=1},
    scaled x ticks=false,
	minor x tick num=1,
	minor y tick num=1,
	xmin=0,
	xmax=17,
    ymax=1,
    ymin=0.01,
	ylabel={Bit Error Rate (BER)},
    axis line style=thick,
    cycle list name=Dark2-6
	] %
    
    
    \addplot+[dashed,line width = 1,mark=square*] table [x=r0, y=HG44, col sep=comma] {summary.csv};
    \addplot+[dashed,line width = 1,mark=*] table [x=r0, y=LG61, col sep=comma] {summary.csv};
    \addplot+[line width = 1,mark=otimes] table [x=r0, y=EGC4461, col sep=comma] {summary.csv};
	
    
	\end{semilogyaxis} %
	\end{tikzpicture}
	
	\caption{\label{fig:ber8} Bit Error Rate of modes with order $N=8$ with varying turbulence strength showing a clear improvement for the diversity case (solid lines).}
\end{figure}

The bit error rate results versus turbulence strength ($r_0$) are shown in Figs.~\ref{fig:ber4} and \ref{fig:ber8}. It is clear that the BER of the EGC case is on average 23\% better than that of the LG SISO case for almost all turbulence strengths. At the weakest turbulence strength, the improvement in diversity BER over both SISO cases is 54\%. Due to the fact that the signal powers were normalised, this can only be due to a diversity gain. 

Interestingly, the HG case is noticeably better than the LG case at medium turbulence strengths. Given that tip/tilt is the dominant aberration in atmospheric turbulence, this result agrees with the findings in Ref.~\cite{ndagano2017} which demonstrate that HG modes are more resilient to tip/tilt aberrations than LG modes. At weaker turbulence strengths, characterised by smaller $D/r_0$ values, the difference between HG and LG is no longer clearly visible. This is expected because the effect of turbulence has become negligible. The diversity gain is still present because of the independent nature of the effect of turbulence on the HG and LG modes.

These results can be put into context by calculating the effective propagation distance gain at a specific BER \cite{Andrews2005}. If Kolmogorov turbulence is assumed, then we can write $r_0$ as a function of $C_n^2$, the refractive index structure parameter, the wavelength, $\lambda$, and the propagation distance, $z$: 

\begin{equation}
r_0=0.185\left(\frac{\lambda^2}{C_n^2 z}\right)^{3/5}
\end{equation}

\noindent This equation can then be solved for $z$, resulting in:

\begin{equation}
\label{eq:distance}
z \approx \frac{0.0600647 \lambda^2}{C_n^2 r_0^{5/3}}
\end{equation}

\noindent In this experiment, $\lambda=660$ nm and a typical value for $C_n^2$ is $10^{-14}$~m$^{-2/3}$ in strong turbulence \cite{Andrews2005}. Three arbitrary BERs were selected and in Table ~\ref{tab:distances} the corresponding $r_0$ values are provided. From Eq.~\ref{eq:distance}, the corresponding theoretical propagation distances are also shown. It is clear that significant improvements in propagation distance are possible using modal diversity.

\begin{table}[ht]
\centering
\begin{tabular}{| l | c c c |} 
 \hline
  Bit Error Rate & 0.04 & 0.08 & 0.15  \\ 
  \hline
  LG$_2^1$ $r_0$ (mm) & 16.6 & 10.2 & 4.5 \\ 
  Diversity $r_0$ (mm) & 12.8 & 7.4 & 2.68  \\ 
  \hline
  LG$_2^1$ Distance (km) & 0.24 & 0.55 & 2.13  \\ 
  Diversity Distance (km) & 0.37 & 0.93 & 5.06  \\ 
  Distance Gain (\%) & 54 & 71 & 137 \\
 \hline
\end{tabular}
\caption{\label{tab:distances} Example relations between $r_0$ and the equivalent propagation distance, $z$, with $C_n^2=10^{-14}$~m$^{-2/3}$ for selected BERs according to Eq.~\ref{eq:distance} for the LG and diversity cases showing a clear improvement for the diversity case. }
\label{table:1}
\end{table}

\medskip
Traditionally, diversity has been theoretically and experimentally demonstrated with a $r_0$ (or greater) separation between transmit and/or receive apertures.  In this work we have outlined the concept of modal diversity, and shown that a significant diversity gain can in fact be achieved in a far more compact system, without $r_0$ separation between apertures or beams.  While we have used orthogonal HG and LG modes of the same order for the demonstration, we expect modal diversity to be observed in any system with judiciously selected mode sets, whether differing in mode order, size or mode type.  We speculate that it should be possible to formalise the concept of ``separate'' in terms of a newly defined modal distance, $r_M$, akin to the previous definition in terms of a physical distance, $r_0$.  That is, how far would two modes have to be separated by in mode space, $r_M$, in order to ensure the same gain as a physical separation of $r_0$?  Importantly, the measured diversity gain is shown to increase the allowable propagation distance (determined by an acceptable error rate) by a significant margin at both strong and weak turbulence strengths. This finding can be used to significantly improve future FSO communication links in terms of range and reliability.

\section*{Funding Information}

The financial assistance of the National Research Foundation (NRF), the Claude Leon Foundation and CONACyT towards this research is acknowledged.

\section*{Acknowledgements}

The authors would like to acknowledge Daniel J. Versfeld for his input during the course of this work.

\bibliographyfullrefs{bib}

\end{document}